\documentclass[11pt]{article}
\usepackage{moriond,epsfig}

\newcommand{\nex}{{\sc neXus}~}
\newcommand{\pt}{$P_{T}$~}

\begin{document}
\vspace*{4cm}
\title{HIGH PT SUPPRESSION WITHOUT JET QUENCHING IN AU+AU COLLISIONS IN NEXUS}

\author{T. PIEROG (H. J. DRESCHER, F.M. LIU, S. OSTAPCHENKO AND K. WERNER)}
\address{FZK, Institut f\"ur Kernphysik, Postfach 3640, 76021 Karlsruhe, GERMANY}

\maketitle\abstracts{Problem of high \pt suppression in RHIC data is generally associated 
to jet quenching in a dense medium. We recently proposed a new approach
to high energy nuclear scattering, which treats the initial stage of heavy
ion collisions in a sophisticated way, and whose numerical solution is
the model {\sc neXus}. Within this model, there is no jet quenching, but the
explicit energy conservation leads to similar results. RHIC high \pt
ratio between AA and pp results can be reproduced for different
centrality bins.}

\section{Introduction \label{intro}}

High \pt  suppression in RHIC Au+Au data is one of the most exciting
results of this experiment~\cite{d'Enterria:2002bw,Adcox:2002pe,Adler:2002xw,Adams:2003kv}.
 Together with back-to-back high \pt hadron correlations~\cite{Adler:2002tq}
 and new results for d+Au collisions~\cite{Adams:2003im}, this suppression appears
 to be an effect of final state interactions. Indeed, jet quenching~\cite{Wang:xy} 
in association with Cronin effect~\cite{Cronin:zm} and nuclear shadowing~\cite{Close:1989ca}
, can describe the data reasonably 
well~\cite{Adams:2003kv}. For a qualitative description it is enough, but for a 
precise quantitative description of high \pt and all the other observables in 
ultra-relativistic heavy ion collisions, we have to take care about other effects. 
Those can lead to significant difference in the energy-loss parameter for instance.

Then it is important to have a proper description of the initial state of this kind 
of interaction. The most sophisticated approach to high energy hadronic interactions is the
so-called Gribov-Regge theory~\cite{Gribov:fc}. This is an effective field
theory, which allows multiple interactions to happen ``in parallel'', with
phenomenological objects called \emph{Pomerons} representing elementary 
interactions~\cite{Baker:cv}. 

We recently presented a new approach~\cite{Werner:ze,Drescher:1999js,Drescher:2000ha,Pierog:2002}, for
hadronic interactions and the initial stage of nuclear collisions, which is
able to solve several of the problems of the Gribov-Regge theory, such as 
a consistent approach  to include both 
soft and hard processes, and the energy conservation both for cross section
and particle production calculations. In both cases, energy is properly
shared between the different interactions happening in parallel. This is the
most important new aspect of our approach, which we consider a first necessary
step to construct a consistent model for high energy nuclear scattering. And this 
leads to interesting results.

\section{\nex \label{nexus}}

\begin{figure}[htbp]
{\par\hfill \psfig{figure=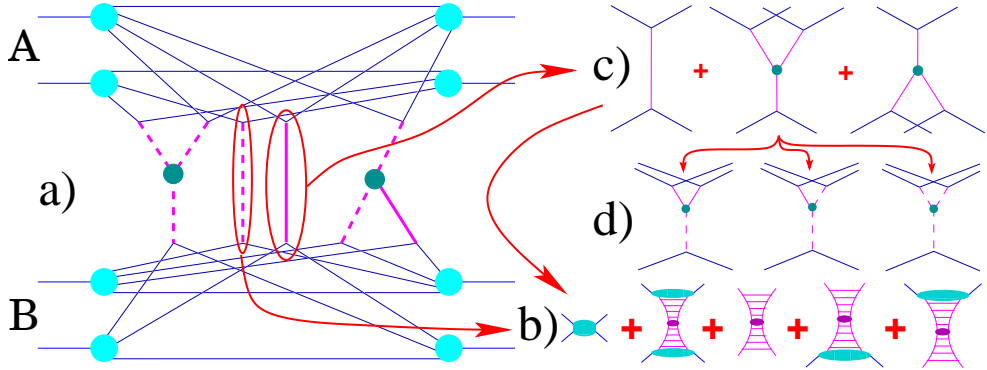,width=13cm} \hfill ~ \par} 
\medskip
\caption{ \label{allin1} a) Multiple elementary interactions (Pomerons)
in \nex. The energy of each proton (blob) is shared between elastic (full
vertical line) and inelastic (dashed vertical line) elementary interactions.
 A Pomeron b) has  soft (blob), hard (ladder)
 and semihard contributions. c) Enhanced diagrams are included and can give
 different inelastic contributions d).}
\end{figure}

We will discuss the basic features of the new approach in a qualitative
fashion. It is an effective theory based on effective elementary interactions.
Multiple interactions happen in parallel in proton-proton and nucleus-nucleus 
collisions.
An elementary interaction is referred to as a Pomeron, and can be either elastic
(uncut Pomeron) or inelastic (cut Pomeron). The spectators
of each proton form remnants, see Fig.~\ref{allin1}a. %

Since a Pomeron is finally identified with two strings, the Pomeron aspect
(to obtain probabilities) and the string aspect
(to obtain particles) are treated in a completely consistent way. \emph{In
both cases energy sharing is considered in a rigorous way, and in
both cases all Pomerons are identical.} To share the energy of the nucleons, we made
a strong and simple assumption that the partition function does not depend of 
the number of elementary interactions. We will discuss this important point 
in the followings.

This theory provides also a consistent treatment for hard and soft
processes: each Pomeron can be expressed in terms of contributions
of different types, soft, hard and semihard ones, cf. Fig.~\ref{allin1}b.
 A hard Pomeron stands
for a hard interaction between valence quarks of initial hadrons. A semihard
one stands for an interaction between sea quarks or gluons but
in which a perturbative process involves in the middle. The high \pt 
particles come from this middle part of the semi-hard (or hard) Pomeron. No perturbative 
process occures at all in soft Pomerons.

A Pomeron is an elementary interaction. 
But those Pomerons may interact with each other at high energy~\cite{Baker:cv,kai86}, 
then they give another type of interaction called 
\emph{enhanced diagram}. There are many types of enhanced diagrams depending on 
the number of Pomerons for each vertices and on the number of vertices. 
In our model, 
effective first order of triple-Pomeron vertices (Y diagrams see
Fig.~\ref{allin1}c) are enough to cure unitarity problem which occure at high energy
without this kind of diagram~\cite{Pierog:2002}. Indeed, Y-type diagrams are screening
corrections which are negative contributions to the cross-section. 
The inelastic contributions (cut enhanced diagrams on Fig.~\ref{allin1}d) of this diagrams
contribute to the increase of the fluctuations in particle production, and in case of
nuclear collision, this type of diagramms correspond to a kind of shadowing 
(modification of the structure function of the nucleons inside a nucleus).

The model \nex is designed to reproduce proton-proton interactions. The initial 
stage of nuclear interaction is obtained by a sophisticated extension of the formalism
 with some approximations for the numeric solution. 
As a consequence, there is neither Cronin Effect nor 
partonic or hadronic final state interaction as jet quenching, hydrodynamic or 
rescattering. Comparison with the data should then be done carrefully.

\section{Results \label{resu}}

Since \nex results for AA are just an extrapolation of pp, but with a proper energy 
sharing scheme, we can compare the high \pt production of Au+Au collisions at 200 Gev
 in \nex with the data, to see what is the effect of the energy-momentum conservation. 
As there is no hadronic final state interactions, the results 
for a $P_{T}<3-4$ GeV should not be regarded as a realistic one. To quantifythe medium effect
, we use the nuclear modification factor $R_{AA}$ defined in eq.~\ref{raa} or
 the ratio $R_{CP}$ of the central yield to the peripheral yield defined in eq.~\ref{rcp}, where
in both cases $<N_{coll}>$ is the mean number of binary collisions for a given 
centrality region in the Glauber model (which does not take into account the 
energy-momentum conservation).

\begin{minipage}[c]{0.5\columnwidth}%
\noindent \begin{center}\begin{equation}
R_{AA}=\frac{\left(1/N_{AA}^{evt}\right)d^{2}N_{AA}/dy dp_{T}}{<N_{coll}>/\sigma _{pp}^{ine}\, \times \, d^{2}\sigma _{pp}/dy /dp_{T}}\label{raa}\end{equation}
\end{center}\end{minipage}%
\hfill
\begin{minipage}[c]{0.45\columnwidth}%
\noindent \begin{center}\begin{equation}
R_{CP}=\frac{<N_{coll}^{peri}>\, \times \, d^{2}N_{cent}/dy dp_{T}}{<N_{coll}^{cent}>\, \times \, d^{2}N_{peri}/dy /dp_{T}}\label{rcp}\end{equation}
\end{center}\end{minipage}%

\begin{figure}[hbtp]
{\par \hfill
\centering \psfig{figure=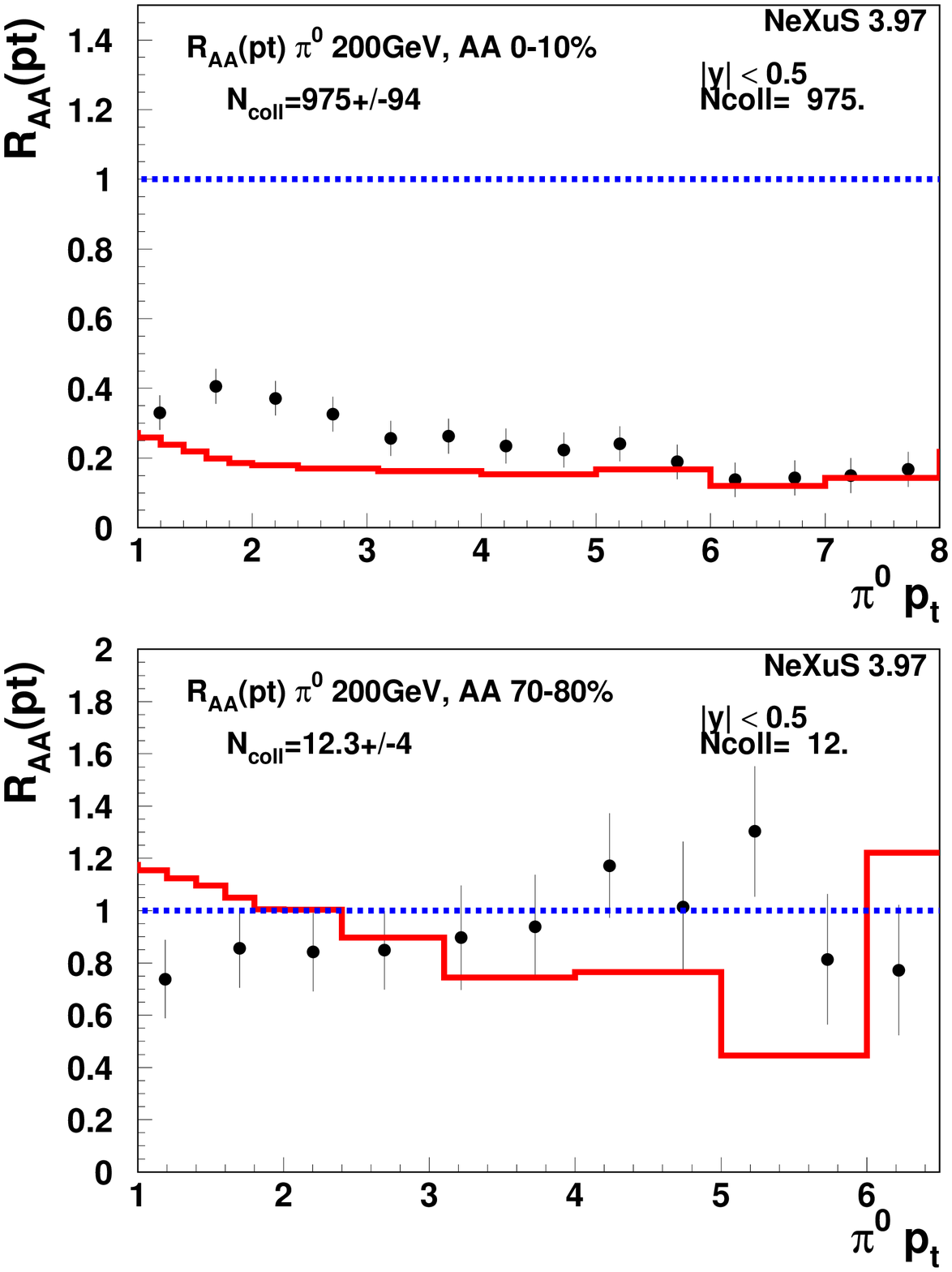,width=7.5cm,height=10cm}\hfill 
\psfig{figure=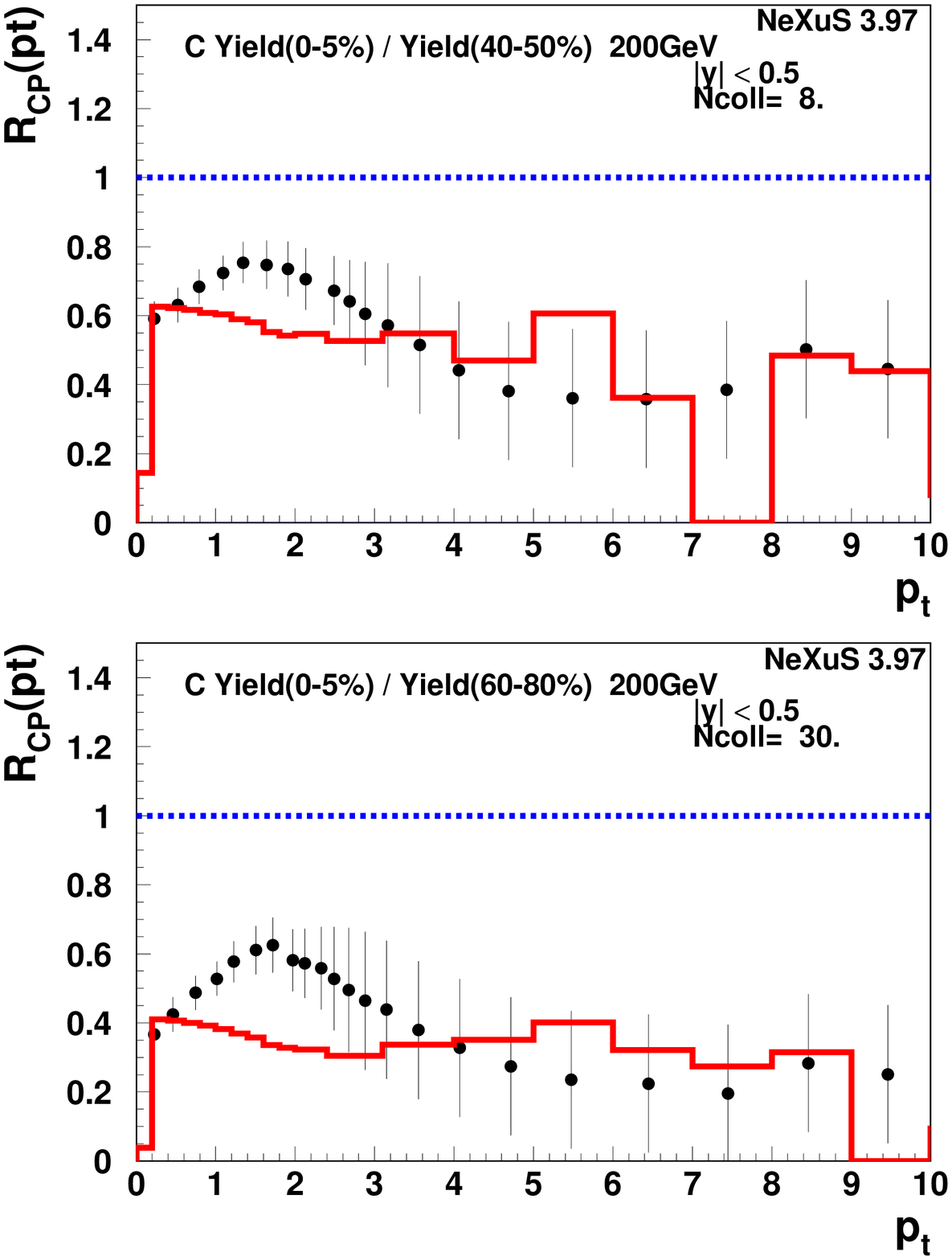,width=7.5cm,height=10cm}\hfill ~
\par}
\medskip
\caption{ \label{results} Left-hand side, nuclear modification factor for $\pi^0$ at 200 GeV 
for two different centrality-selected spectra. Points are experimental data from 
Phoenix collaboration$^{17}$. Right-hand side, ratio of central 
to peripheral charged particles yield for two different peripheral 
centrality-selected spectra. Points are experimental data from 
STAR collaboration$^{18}$.}
\end{figure}

In fig.~\ref{results}, the experimental ratio $R_{AA}$ for $\pi^0$ is compared
 to \nex predictions for the 0-10\% central events (top) and 70-80\% peripheral events (bottom 
left-hand side), together with the $R_{CP}$ for charged hadrons ($(h^++h^-)/2$) where 
central means 0-5\% central events and peripheral means 40-50\% (top) or 60-80\% 
peripheral events (bottom right-hand size).

We can see that in all cases, an energy-momentum conservation scheme at the level of
the cross-section calculation (which fixes the number of binary collisions) can lead to
a suppression of the high \pt produced particles which is compatible with the data.

\section{Discussion}

Of course, we are not claiming that this kind of process explains the high \pt suppression.
The recent d+Au data do not show any particular suppression for high \pt, while in \nex a 
suppression appear. A Cronin effect would explain a part of the difference, but surely not
all. In fact in our scheme, the suppression due to the energy conservation mechanism 
is maximal because
of our simple choice for the partition function. It has to be seen as a maximum limit
of this effect. A better understanding of this part
of our formalism, which can be done partially with pp data, should lead to a weaker
suppression. Together with a strong Cronin effect, it could appear like a weak Cronin 
effect in d+Au reaction. For heavy ion reactions, this can change the value of the needed
energy-loss.
Thus we want to emphasize that for a quantitative description of heavy ion collision data,
and for a real comprehension of the complex processes involved in this kind of reaction, 
it is important to take care about energy-momentum conservation. It can play a 
non-negligeable role to fix the proportion of all the other processes like the Cronin
effect or the jet quenching.

S. O. was supported by the German Ministry for Education and Research (BMBF,
grant 05 CU1VK1/9) and F.M. L. is supported by the German Alexander von Humboldt Foundation.

\end{document}